\documentclass[reprint,amsmath,amssymb,aps]{revtex4-2}

\usepackage{graphicx}
\usepackage[hidelinks]{hyperref}
\usepackage{caption}
\usepackage{subcaption}

\begin{document}

\title{New asymptotically flat gravitational instanton}

\author{Edward Teo}
\affiliation{Department of Physics, National University of Singapore, Singapore 117551, Singapore}%

\begin{abstract}
A new two-parameter asymptotically flat (AF) toric gravitational instanton is identified as a special case of the Euclidean double Kerr-NUT solution, by imposing certain symmetry and regularity conditions on its rod structure. These conditions are solved explicitly, except for one which takes the form of a fifth-order polynomial equation. This gravitational instanton has Euler number $\chi=4$ and Hirzebruch signature $\tau=0$, and its global topology is $\mathbb{C}P^2\#\overline{\mathbb{C}P^2}$ with a circle $S^1$ removed appropriately. It is the third of an infinite sequence of AF toric gravitational instantons that was proved to exist by Li and Sun, the first two being the Kerr and Chen--Teo instantons. It is also the first known example of an AF gravitational instanton that is not Hermitian.
\end{abstract}

\maketitle

Gravitational instantons are complete non-singular positive-definite solutions to the Einstein equations \cite{Hawking:1976,Gibbons:1979}. In the case when the cosmological constant  vanishes, they are non-compact Ricci-flat Riemannian 4-manifolds and can be classified into one of several classes depending on their asymptotic behaviour. This includes the class of asymptotically flat (AF) gravitational instantons \cite{Gibbons:1979a}, which have a cubic volume growth and a boundary at infinity that is $S^1\times S^2$.

The study of gravitational instantons first took off in the late 1970's, driven by their importance in the Euclidean path-integral approach to quantum gravity \cite{Gibbons:1993}. In this approach, the path integral is performed over all metrics with Euclidean signature, subject to fixed boundary conditions. Since path integrals are in general impossible to evaluate exactly, an approximation has to be made. In the stationary-phase approximation, the path integral is dominated by classical solutions to the Einstein equations, i.e., gravitational instantons. 

AF gravitational instantons in particular contribute to the partition function in the grand canonical ensemble at a fixed temperature and angular velocity \cite{Gibbons:1977}. In this case, the path integral is performed over all Euclidean metrics obeying certain periodicity conditions on the coordinates at infinity. It turns out that the Euclidean Kerr instanton satisfies these precise conditions, and so will contribute to the partition function.

Back in the 1970's, the Euclidean Kerr instanton (with Euclidean Schwarzschild as a special case) was the only non-trivial AF gravitational instanton known. This led to the conjecture \cite{Lapedes:1980}---essentially the Euclidean analogue of the black-hole uniqueness theorem---that there are no other AF gravitational instantons. However, this conjecture was shown to be false in 2011 with the discovery of a new AF gravitational instanton by Chen and the present author \cite{Chen:2011}. This has sparked a resurgence of interest in the study of AF and other gravitational instantons in the past several years (e.g., \cite{Kunduri:2021,Aksteiner:2021,Biquard:2021,Nilsson:2023,Aksteiner:2023,Dunajski:2024,Li:2025,Araneda:2025}).

A gravitational instanton is toric if it has a ${\cal T}=U(1)\times U(1)$ isometry. It turns out that many of the known gravitational instantons are toric, including the Chen--Teo instanton. Toric gravitational instantons can be studied and classified using the rod-structure formalism of \cite{Chen:2010}, which characterises the fixed points of the ${\cal T}$ isometry. The rod structure of a given gravitational instanton consists of a sequence of so-called rods, separated by points known as turning points. Turning points are fixed points of the whole ${\cal T}$ isometry group, while rods remain invariant under a one-dimensional subgroup of ${\cal T}$. A two-dimensional vector can be associated to each rod, which generates the particular subgroup it is invariant under.

Recently, Li and Sun \cite{Li:2025} proved that the Kerr and Chen--Teo instantons are just the first two members of an infinite family of AF toric gravitational instantons labelled by a positive integer $n$. The rod structure was found for each $n$, and it consists of $n+1$ turning points. The rod structures for the Kerr and Chen--Teo instantons are recovered when $n=1$ and 2, respectively. Since the proof in \cite{Li:2025} is non-constructive, the explicit metrics of the AF gravitational instantons when $n\geq3$ remained unknown. The purpose of this letter is to construct the metric for the $n=3$ case, i.e., a new AF toric gravitational instanton with four turning points.

Our starting point is the Euclidean double Kerr-NUT solution. Although this solution can be obtained by an analytic continuation of the corresponding Lorentzian solution \cite{Kramer:1980}, it can also be directly constructed by applying the inverse-scattering method of Belinski and Zakharov \cite{Belinski:2001} on an appropriate seed. One way, as described in \cite{Chen:2015}, is by performing a 4-soliton transformation on Euclidean flat space with so-called BZ parameters $C_k$ ($k=1,\dots,4$). The resulting solution can be written in Weyl--Papapetrou coordinates as
\begin{align}
\label{metric}
{\rm d}s^2&=\frac{\mu_1\mu_2\mu_3\mu_4F}{H}({\rm d}\psi+\omega\,{\rm d}\phi)^2+\frac{\rho^2H}{\mu_1\mu_2\mu_3\mu_4F}\,{\rm d}\phi^2\cr
&\quad+\frac{k_0H}{(\mu_1\mu_2\mu_3\mu_4)^3R_{11}R_{22}R_{33}R_{44}}({\rm d}\rho^2+{\rm d}z^2)\,,
\end{align}
where we have defined
\begin{align}
\mu_k\equiv\sqrt{\rho^2+(z-z_k)^2}-(z-z_k)\,,\qquad R_{kl}\equiv\rho^2+\mu_k\mu_l\,.
\end{align}
If we further define $\mu_{kl}\equiv\mu_k-\mu_l$, the functions $H$, $F$ and $\omega$, after some simplification, are given by
\begin{widetext}
\begin{subequations}
\begin{align}
H&=\big[\rho^2\mu_{12}\mu_{13}\mu_{14}\mu_{23}\mu_{24}\mu_{34}(\rho^4+C_1C_2C_3C_4\mu_1\mu_2\mu_3\mu_4)
+R_{13}R_{14}R_{23}R_{24}\mu_{12}\mu_{34}(C_1C_2\mu_1\mu_2+C_3C_4\mu_3\mu_4)\cr
&\quad+R_{12}R_{13}R_{24}R_{34}\mu_{14}\mu_{23}(C_1C_4\mu_1\mu_4+C_2C_3\mu_2\mu_3)-R_{12}R_{14}R_{23}R_{34}\mu_{13}\mu_{24}(C_1C_3\mu_1\mu_3+C_2C_4\mu_2\mu_4)\big]^2\cr
&\quad
-\big[R_{12}R_{13}R_{14}\mu_{23}\mu_{24}\mu_{34}(\rho^2C_1\mu_1-C_2C_3C_4\mu_2\mu_3\mu_4)-R_{12}R_{23}R_{24}\mu_{13}\mu_{14}\mu_{34}(\rho^2C_2\mu_2-C_1C_3C_4\mu_1\mu_3\mu_4)\cr
&\quad+R_{13}R_{23}R_{34}\mu_{12}\mu_{14}\mu_{24}(\rho^2C_3\mu_3-C_1C_2C_4\mu_1\mu_2\mu_4)-R_{14}R_{24}R_{34}\mu_{12}\mu_{13}\mu_{23}(\rho^2C_4\mu_4-C_1C_2C_3\mu_1\mu_2\mu_3)\big]^2,\\
F&=\big[\rho^4(1+C_1C_2C_3C_4)\mu_{12}\mu_{13}\mu_{14}\mu_{23}\mu_{24}\mu_{34}
-(C_1C_2+C_3C_4)R_{13}R_{14}R_{23}R_{24}\mu_{12}\mu_{34}\cr
&\quad+(C_1C_3+C_2C_4)R_{12}R_{14}R_{23}R_{34}\mu_{13}\mu_{24}
-(C_1C_4+C_2C_3)R_{12}R_{13}R_{24}R_{34}\mu_{14}\mu_{23}\big]^2\cr
&\quad+\rho^2\big[(C_1+C_2C_3C_4)R_{12}R_{13}R_{14}\mu_{23}\mu_{24}\mu_{34}-(C_2+C_1C_3C_4)R_{12}R_{23}R_{24}\mu_{13}\mu_{14}\mu_{34}\cr
&\quad+(C_3+C_1C_2C_4)R_{13}R_{23}R_{34}\mu_{12}\mu_{14}\mu_{24}-(C_4+C_1C_2C_3)R_{14}R_{24}R_{34}\mu_{12}\mu_{13}\mu_{23}\big]^2,\\
\omega&=\frac{1}{\mu_1\mu_2\mu_3\mu_4F}
\Big\{2C_1C_2C_3R_{11}R_{14}R_{22}R_{24}R_{33}R_{34}\mu_{14}\mu_{24}\mu_{34}(\rho^2+C_4^2\mu_4^2)\cr
&\quad\times\big[R_{12}\mu_1\mu_3^2\mu_{12}(\rho^2-\mu_2^2)-R_{13}\mu_2^2\mu_3\mu_{13}(\rho^2-\mu_1^2)+R_{23}\mu_1^2\mu_2\mu_{23}(\rho^2-\mu_3^2)\big]\cr
&\quad
+2C_1C_2C_4R_{11}R_{13}R_{22}R_{23}R_{34}R_{44}\mu_{13}\mu_{23}\mu_{34}(\rho^2+C_3^2\mu_3^2)\cr
&\quad\times\big[R_{12}\mu_2\mu_4^2\mu_{12}(\rho^2-\mu_1^2)+R_{24}\mu_1^2\mu_4\mu_{24}(\rho^2-\mu_2^2)-R_{14}\mu_1\mu_2^2\mu_{14}(\rho^2-\mu_4^2)\big]\cr
&\quad
+2C_1C_3C_4R_{11}R_{12}R_{23}R_{24}R_{33}R_{44}\mu_{12}\mu_{23}\mu_{24}(\rho^2+C_2^2\mu_2^2)\cr
&\quad\times\big[R_{13}\mu_1\mu_4^2\mu_{13}(\rho^2-\mu_3^2)-R_{14}\mu_3^2\mu_4\mu_{14}(\rho^2-\mu_1^2)+R_{34}\mu_1^2\mu_3\mu_{34}(\rho^2-\mu_4^2)\big]\cr
&\quad
-2C_2C_3C_4R_{12}R_{13}R_{14}R_{22}R_{33}R_{44}\mu_{12}\mu_{13}\mu_{14}(\rho^2+C_1^2\mu_1^2)\cr
&\quad\times\big[R_{23}\mu_2\mu_4^2\mu_{23}(\rho^2-\mu_3^2)-R_{24}\mu_3^2\mu_4\mu_{24}(\rho^2-\mu_2^2)+R_{34}\mu_2^2\mu_3\mu_{34}(\rho^2-\mu_4^2)\big]\cr
&\quad-C_1R_{11}R_{12}R_{13}R_{14}\mu_{12}\mu_{13}\mu_{14}\big[
\rho^2\mu_{23}^2\mu_{24}^2\mu_{34}^2(\rho^6+C_2^2C_3^2C_4^2\mu_2^2\mu_3^2\mu_4^2)+R_{23}^2R_{24}^2\mu_{34}^2(\rho^2C_2^2\mu_2^2+C_3^2C_4^2\mu_3^2\mu_4^2)\cr
&\quad+R_{23}^2R_{34}^2\mu_{24}^2(\rho^2C_3^2\mu_3^2+C_2^2C_4^2\mu_2^2\mu_4^2)+R_{24}^2R_{34}^2\mu_{23}^2(\rho^2C_4^2\mu_4^2+C_2^2C_3^2\mu_2^2\mu_3^2)\big]\cr
&\quad+C_2R_{12}R_{22}R_{23}R_{24}\mu_{12}\mu_{23}\mu_{24}\big[
\rho^2\mu_{13}^2\mu_{14}^2\mu_{34}^2(\rho^6+C_1^2C_3^2C_4^2\mu_1^2\mu_3^2\mu_4^2)
+R_{13}^2R_{14}^2\mu_{34}^2(\rho^2C_1^2\mu_1^2+C_3^2C_4^2\mu_3^2\mu_4^2)\cr
&\quad+R_{13}^2R_{34}^2\mu_{14}^2(\rho^2C_3^2\mu_3^2+C_1^2C_4^2\mu_1^2\mu_4^2)+R_{14}^2R_{34}^2\mu_{13}^2(\rho^2C_4^2\mu_4^2+C_1^2C_3^2\mu_1^2\mu_3^2)\big]\cr
&\quad-C_3R_{13}R_{23}R_{33}R_{34}\mu_{13}\mu_{23}\mu_{34}\big[
\rho^2\mu_{12}^2\mu_{14}^2\mu_{24}^2(\rho^6+C_1^2C_2^2C_4^2\mu_1^2\mu_2^2\mu_4^2)+R_{12}^2R_{14}^2\mu_{24}^2(\rho^2C_1^2\mu_1^2+C_2^2C_4^2\mu_2^2\mu_4^2)\cr
&\quad+R_{12}^2R_{24}^2\mu_{14}^2(\rho^2C_2^2\mu_2^2+C_1^2C_4^2\mu_1^2\mu_4^2)+R_{14}^2R_{24}^2\mu_{12}^2(\rho^2C_4^2\mu_4^2+C_1^2C_2^2\mu_1^2\mu_2^2)\big]\cr
&\quad+C_4R_{14}R_{24}R_{34}R_{44}\mu_{14}\mu_{24}\mu_{34}\big[\rho^2\mu_{12}^2\mu_{13}^2\mu_{23}^2(\rho^6+C_1^2C_2^2C_3^2\mu_1^2\mu_2^2\mu_3^2)
+R_{12}^2R_{13}^2\mu_{23}^2(\rho^2C_1^2\mu_1^2+C_2^2C_3^2\mu_2^2\mu_3^2)\cr
&\quad+R_{12}^2R_{23}^2\mu_{13}^2(\rho^2C_2^2\mu_2^2+C_1^2C_3^2\mu_1^2\mu_3^2)+R_{13}^2R_{23}^2\mu_{12}^2(\rho^2C_3^2\mu_3^2+C_1^2C_2^2\mu_1^2\mu_2^2)\big]
\Big\}\,.
\end{align}
\end{subequations}
\end{widetext}
There are eight parameters contained in this solution: $z_k$, $C_k$, for $k=1,\dots,4$. However, only seven of them are physical as one of the $z_k$'s can be arbitrarily fixed by a translation in the $z$-direction. The constant $k_0$ in (\ref{metric}) can also be arbitrarily set to any positive constant, but it is useful to retain it.

\begin{figure}[h]
      \centering
      \includegraphics[scale=0.83]{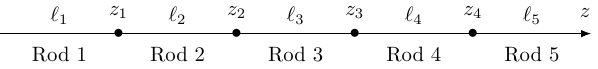}
  \caption{The rod structure of the Euclidean double Kerr-NUT solution (\ref{metric}), with turning points $z_1,\dots,z_4$, and rod vectors $\ell_1,\dots,\ell_5$ given by (\ref{rod vectors}).}
      \label{rod}
\end{figure}

The Euclidean double Kerr-NUT solution admits two commuting Killing vector fields $\frac{\partial}{\partial \psi}$ and $\frac{\partial}{\partial \phi}$, whose fixed points can be analysed using the rod-structure formalism of \cite{Chen:2010}. Its rod structure consists of five rods, lying along the $z$-axis and separated by the four turning points $z_k$. The direction of each rod has the form $a\frac{\partial}{\partial \psi}+b\frac{\partial}{\partial \phi}$, which will be written as $(a,b)$ for brevity. Without loss of generality, we assume the ordering $z_1<z_2<z_3<z_4$, so that the rods lie along the intervals $(-\infty,z_1]$, $[z_1,z_2]$, $[z_2,z_3]$, $[z_3,z_4]$, $[z_4,\infty)$. We label these five rods in order from left to right, as in Fig.~\ref{rod}. If we define $z_{ij}\equiv z_i-z_j$, the corresponding rod vectors are \cite{Chen:2015}
\begin{widetext}
\begin{subequations}
\label{rod vectors}
\begin{align}
\ell_1&=\big({-}8\sqrt{k_0}(C_1C_2C_3z_{12}z_{13}z_{23}-C_1C_2C_4z_{12}z_{14}z_{24}+C_1C_3C_4z_{13}z_{14}z_{34}-C_2C_3C_4z_{23}z_{24}z_{34}),\cr
&\qquad -4\sqrt{k_0}(C_1C_2z_{12}z_{34}-C_1C_3z_{13}z_{24}+C_1C_4z_{14}z_{23}+C_2C_3z_{14}z_{23}-C_2C_4z_{13}z_{24}+C_3C_4z_{12}z_{34})\big)\,,\\
\ell_2&=\big(8\sqrt{k_0}(C_2C_3z_{12}z_{13}z_{23}-C_2C_4z_{12}z_{14}z_{24}+C_3C_4z_{13}z_{14}z_{34}-C_1C_2C_3C_4z_{23}z_{24}z_{34}),\cr
&\qquad 4\sqrt{k_0}(C_2z_{12}z_{34}-C_3z_{13}z_{24}+C_4z_{14}z_{23}+C_1C_2C_3z_{14}z_{23}-C_1C_2C_4z_{13}z_{24}+C_1C_3C_4z_{12}z_{34})\big)\,,\\
\ell_3&=\big(8\sqrt{k_0}(C_3z_{12}z_{13}z_{23}-C_4z_{12}z_{14}z_{24}-C_1C_3C_4z_{23}z_{24}z_{34}+C_2C_3C_4z_{13}z_{14}z_{34}),\cr
&\qquad 4\sqrt{k_0}(z_{12}z_{34}+C_1C_3z_{14}z_{23}-C_1C_4z_{13}z_{24}-C_2C_3z_{13}z_{24}+C_2C_4z_{14}z_{23}+C_1C_2C_3C_4z_{12}z_{34})\big)\,,\\
\ell_4&=\big({-}8\sqrt{k_0}(z_{12}z_{13}z_{23}-C_1C_4z_{23}z_{24}z_{34}+C_2C_4z_{13}z_{14}z_{34}-C_3C_4z_{12}z_{14}z_{24}),\cr
&\qquad -4\sqrt{k_0}(C_1z_{14}z_{23}-C_2z_{13}z_{24}+C_3z_{12}z_{34}+C_1C_2C_4z_{12}z_{34}-C_1C_3C_4z_{13}z_{24}+C_2C_3C_4z_{14}z_{23})\big)\,,\\
\ell_5&=\big(8\sqrt{k_0}(C_1z_{23}z_{24}z_{34}-C_2z_{13}z_{14}z_{34}+C_3z_{12}z_{14}z_{24}-C_4z_{12}z_{13}z_{23}),\cr
&\qquad -4\sqrt{k_0}(C_1C_2z_{12}z_{34}-C_1C_3z_{13}z_{24}+C_1C_4z_{14}z_{23}+C_2C_3z_{14}z_{23}-C_2C_4z_{13}z_{24}+C_3C_4z_{12}z_{34})\big)\,.
\end{align}
\end{subequations}
\end{widetext}
Note that the second components of $\ell_1$ and $\ell_5$ are equal.

The set of rod vectors (\ref{rod vectors}) can also be obtained from those of the Lorentzian double Kerr-NUT solution by an appropriate analytic continuation \cite{Chen:2015}. Recall that in that case (e.g., \cite{Herdeiro:2008}), Rods 2 and 4 are the horizons of the two black holes, and their surface gravities and angular velocities are determined by the directions of these rods. Also recall that the overall NUT charge of the space-time is determined by the relative directions of Rods 1 and 5. A formally similar situation applies for the Euclidean solution.

Now to obtain an AF solution from the Euclidean double Kerr-NUT solution, we first set the overall NUT charge to zero by requiring that $\ell_1=\ell_5$. We then set the (Euclidean) surface gravities and angular velocities of the two horizons to be equal by requiring that $\ell_2=\ell_4$. It turns out that these two conditions can be simultaneously satisfied if
\begin{subequations}
\label{condition 12}
\begin{gather}
\label{condition 1}
z_3=-z_2\,,\qquad z_4=-z_1\,,\\
C_3=-\frac{1}{C_2}\,,\qquad C_4=-\frac{1}{C_1}\,.
\end{gather}
\end{subequations}
In particular, (\ref{condition 1}) implies that Rods 2 and 4 have the same length, and are equidistant from the origin $z=0$.

If we centre ourselves on Rod 2 and push Rod 4 to infinity, the double Kerr-NUT solution will reduce to a single Kerr-NUT solution, with rod vectors $\ell_1$, $\ell_2$ and $\ell_3$. Assuming these three vectors are linearly independent, the only completely regular solution contained within it was shown in \cite{Chen:2010} to be the Taub-bolt instanton \cite{Page:1978}. Recall that the rod vectors for this instanton satisfy \cite{Chen:2010}
\begin{align}
\label{rod condition}
\ell_1\pm\ell_2=\ell_3\,.
\end{align}
We thus further impose this condition on the double Kerr-NUT solution, as a necessary condition for it to be regular. Since $\ell_1=\ell_5$ and $\ell_2=\ell_4$, (\ref{rod condition}) also implies that $\ell_3\mp\ell_4=\ell_5$. This means if we centre ourselves on Rod 4 and push Rod 2 to infinity, we again obtain a Taub-bolt instanton, but with opposite NUT charge to the first one. Hence our solution will contain two oppositely charged Taub-bolt instantons as special limits.

\begin{figure}[t]
    \begin{subfigure}[b]{0.5\textwidth}
      \centering
      \includegraphics[scale=0.68]{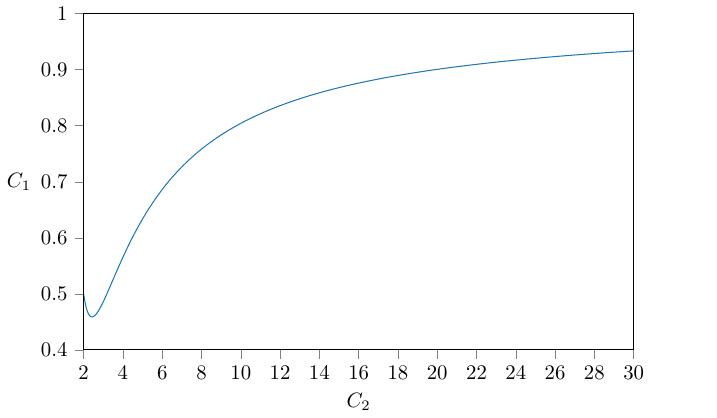}
      \caption{}
    \end{subfigure}
    \hfill
    \begin{subfigure}[b]{0.5\textwidth}
      \centering
	  \includegraphics[scale=0.68]{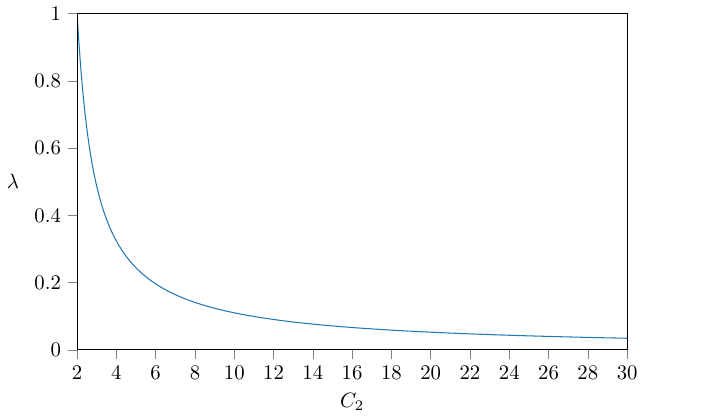}
      \caption{}
    \end{subfigure}
  \caption{Plot of (a) $C_1$ vs.~$C_2$; (b) $\lambda$ vs.~$C_2$ for the relevant set of solutions to (\ref{P}).}
      \label{graphs}
\end{figure}

Without loss of generality, we take the positive sign in (\ref{rod condition}). Unfortunately, this equation does not have an explicit solution. It can be written as
\begin{align}
\label{condition 3}
z_2=\lambda z_1\,,\qquad\lambda\equiv\frac{2C_2+(C_1-1)(C_2^2+1)}{2C_1+(C_1-1)^2C_2}\,,
\end{align}
where $C_1$ and $C_2$ satisfy the constraint
\begin{widetext}
\begin{align}
\label{P}
&(C_1-1)^3C_2^6-2(C_1-1)(C_1^3-C_1^2-3C_1+1)C_2^5+C_1(C_1^4-C_1^3-11C_1^2+9C_1+10)C_2^4+4(C_1^4-4C_1^2-1)C_2^3\cr
&\quad+C_1(C_1-1)(C_1^3+5C_1+10)C_2^2+2(C_1^4+4C_1^2+2C_1+1)C_2+C_1^3-7C_1^2-C_1-1=0\,.
\end{align}
\end{widetext}
Note that (\ref{P}) is a fifth-order polynomial equation in $C_1$ and a sixth-order one in $C_2$ (although it is somewhat shorter in the latter form). The set of solutions that is relevant for us is the one in which $C_2>2$. Fig.~\ref{graphs} shows how $C_1$ and $\lambda$ depend on $C_2$. It can be seen that $C_1\rightarrow\frac{1}{2}$ 
and $\lambda\rightarrow1$ in the limit $C_2\rightarrow2$, and that $C_1\rightarrow1$ and $\lambda\rightarrow0$ in the limit $C_2\rightarrow\infty$. We note that $C_1$ has a minimum of $\simeq0.45884$ at $C_2\simeq2.42775$, and is bounded by 1 from above. We also note that $0<\lambda<1$, as it should since $z_1<z_2<0$.

To confirm that this set of solutions contains the two oppositely charged Taub-bolt instantons, we first centre ourselves on Rod 2 by an appropriate translation of the $z$ coordinate. We then take the limits $C_2\rightarrow2$ and $z_1\rightarrow-\infty$, such that $z_{12}$ remains finite. This effectively pushes Rod 4 to infinity. In this limit, it can be checked that the metric (\ref{metric})---with an appropriate rescaling of $k_0$ and all the above conditions imposed---reduces to that of a Taub-bolt instanton with negative NUT charge, in the form of Eq.~(5.13) in \cite{Chen:2015} with $C_1=\frac{1}{2}$ and $C_2=2$. An analogous procedure can be performed on Rod 4, to obtain a Taub-bolt instanton with the opposite NUT charge.

We briefly turn to the other limit $C_2\rightarrow\infty$. Since $\lambda\rightarrow0$, the length of Rod 3 shrinks to zero in this limit. At the same time, the fact that $C_1\rightarrow1$ means Rod 1 joins up with Rod 2 \cite{Chen:2015}. It can be checked that a flat space is in fact obtained. 

To summarise, the new AF gravitational instanton is obtained from the Euclidean double Kerr-NUT solution (\ref{metric}) by first imposing the conditions (\ref{condition 12}) and (\ref{condition 3}). The condition (\ref{P}) will then have to be (numerically) solved for, to obtain $C_1$ in terms of $C_2$. The result is a two-parameter family of solutions, expressed in terms of $z_1$ and $C_2$. Since $|z_1|$ is a scaling parameter, there is just one non-trivial parameter $C_2$.

To ensure that the solution is free of conical and orbifold singularities, we need to make appropriate identifications on the coordinates $\psi$ and $\phi$. If we write $\ell_1=(a,b)$ and $\ell_2=(c,d)$, the identifications
\begin{align}
(\psi,\phi)&\rightarrow\left(\psi+2\pi a,\phi+2\pi b\right),\cr (\psi,\phi)&\rightarrow\left(\psi+2\pi c,\phi+2\pi d\right),
\end{align}
have to be made to avoid conical singularities along these two rods, and an orbifold singularity at the turning point between the rods. Since every adjacent pair of rod vectors $\{\ell_k,\ell_{k+1}\}$ is related by a $GL(2,\mathbb{Z})$ transformation to every other adjacent pair, there are no conical singularities along any of the rods nor orbifold singularities at any of the turning points \cite{Chen:2010}.

It is possible to bring the rod structure to standard orientation \cite{Chen:2010}, by performing the following coordinate transformation:
\begin{gather}
\psi=c\tilde\psi+a\tilde\phi\,,\qquad
\phi=d\tilde\psi+b\tilde\phi\,,\cr
\rho=\frac{1}{|ad-bc|}\,\tilde\rho\,,\qquad
z=\frac{1}{|ad-bc|}\,\tilde z\,.
\end{gather}
In the new coordinates $(\tilde\psi,\tilde\phi)$, the set of rod vectors will take the form $\tilde\ell_1=(0,1)$, $\tilde\ell_2=(1,0)$, $\tilde\ell_3=(1,1)$, $\tilde\ell_4=(1,0)$ and $\tilde\ell_5=(0,1)$, which are consistent with the results in \cite{Li:2025}. The identifications needed to ensure absence of conical and orbifold singularities are then simply
\begin{align}
(\tilde\psi,\tilde\phi)\rightarrow(\tilde\psi,\tilde\phi+2\pi)\,,\qquad (\tilde\psi,\tilde\phi)\rightarrow(\tilde\psi+2\pi,\tilde\phi)\,.
\end{align}
However, since the metric components are rather complicated in these new coordinates, we will not perform this coordinate transformation explicitly.

The direction pair $\{\tilde\ell_1,\tilde\ell_2\}$ can be identified as the pair of independent $2\pi$-periodic generators of the isometry group $\cal T$ of this gravitational instanton. Note that $\frac{\partial}{\partial\tilde\psi}$ generates an $S^1$ which blows up at infinity. However, if $\frac{d}{b}$ is a rational number, $\frac{\partial}{\partial\psi}$ generates closed and finite orbits at infinity. This is similar to the case of the Kerr \cite{Chen:2010} and Chen--Teo \cite{Chen:2011} instantons.

To confirm that the solution is free of curvature singularities, we should check that the $g_{\rho\rho}$ component of the metric is positive everywhere in the half plane $(\rho\geq0,-\infty<z<\infty)$. Since its form is sufficiently complicated, we have resorted to checking this numerically. Indeed, we have verified that $g_{\rho\rho}$ is positive for a large sampling of parameter values in the allowed range of $C_2$. We have also similarly verified that the $g_{\psi\psi}$ component of the metric is positive everywhere in the half-plane, except at the turning points where it is zero. It follows that the signature of the solution is indeed positive-definite.

Since the directions of Rods 1 and 5 are identical for any AF gravitational instanton, they can be joined up to a single rod at infinity by adding a manifold $S^1\times\mathbb{R}^3$ to the gravitational instanton. In the present case, the rod structure of $\mathbb{C}P^2\#\overline{\mathbb{C}P^2}$ is obtained. This gravitational instanton was found by Page \cite{Page:1978a}, as a solution to the Einstein equations with a positive cosmological constant. It follows that the global topology of the new AF gravitational instanton is $\mathbb{C}P^2\#\overline{\mathbb{C}P^2}$ with a circle appropriately removed, i.e., $\mathbb{C}P^2\#\overline{\mathbb{C}P^2}\,\backslash\, S^1$. It has Euler number $\chi=4$ and Hirzebruch signature $\tau=0$, which are the same as those for the Page instanton. Recall that the Euler number is just the number of turning points in the rod structure, while the Hirzebruch signature can be calculated using say the method of \cite{Nilsson:2023}.

Until recently, all known examples of Ricci-flat gravitational instantons---including the Chen--Teo instanton---were Hermitian \cite{Aksteiner:2021}. It was conjectured in \cite{Aksteiner:2023} that all AF gravitational instantons are Hermitian. However, this was shown to be false by Li and Sun \cite{Li:2025}, for the family of AF toric gravitational instantons they found when the number of turning points is more than three. The new gravitational instanton is therefore the first known example of an AF gravitational instanton that is not Hermitian.

The construction of the present solution using the inverse scattering method, as well as its previous use to construct the Chen--Teo instanton \cite{Chen:2011}, suggests that this method can be used to construct AF toric gravitational instantons with more than four turning points. If the number of turning points $n$ is even, we conjecture that the corresponding gravitational instanton is a special case of the $n$-soliton solution of \cite{Chen:2015}. If $n$ is odd, we conjecture that the corresponding gravitational instanton can be obtained from the $(n+1)$-soliton solution by eliminating one of the solitons in the way described in \cite{Chen:2015}. It would be interesting to try to construct these gravitational instantons explicitly, although their forms would likely be very complicated.

\bigskip
\noindent
\textit{Acknowledgements}. The author is grateful to Yu Chen for valuable discussions.

\end{document}